# Spin-dependent Seebeck Effect in Aharonov-Bohm Rings with Rashba and Dresselhaus Spin-orbit Interactions

Bin Liu,[a] Yunyun Li,[a] Jun Zhou,[a,*] Tsuneyoshi Nakayama,[a,b]
and Baowen Li[c]

[a] Center for Phononics and Thermal Energy Science, School of Physics Science and Engineering, Tongji University, Shanghai 200092, People's Republic of China
[b] Hokkaido University, Sapporo 060-0826, Japan
[c] Department of Mechanical Engineering, University of Colorado, Boulder 80309, USA

**ABSTRACT:** We theoretically investigate the spin-dependent Seebeck effect in an Aharonov-Bohm mesoscopic ring in the presence of both Rashba and Dresselhaus spin-orbit interactions under magnetic flux perpendicular to the ring. We apply the Green's function method to calculate the spin Seebeck coefficient employing the tight-binding Hamiltonian. It is found that the spin Seebeck coefficient is proportional to the slope of the energy-dependent transmission coefficients. We study the strong dependence of spin Seebeck coefficient on the Fermi energy, magnetic flux, strength of spin-orbit coupling, and temperature. Maximum spin Seebeck coefficients can be obtained when the strengths of Rashba and Dresselhaus spin-orbit couplings are slightly different. The spin Seebeck coefficient can be reduced by increasing temperature and disorder.

*Keywords:* Spin-orbit interaction, Spin-dependent Seebeck effect, Aharonov-Bohm Effect, mesoscopic ring

PACS number: 71.70.Ej, 73.23.-b, 73.50.Lw

## 1. Introduction

In the last decades, enormous efforts have been devoted to utilize the spins of electrons in mesoscopic devices, which are usually referred to spintronics [1]. One of the major goals of the spintronics is the generation of spin polarized currents, preferably in semiconductor systems [2]. Recently, spin current generation due to temperature gradient, which is called the spin Seebeck effect (SSE), has been experimentally observed in magnetic materials [3,4,5,6,7] in which the spins are carried by magnons. Later on, the SSE has been also observed in nonmagnetic materials in the presence of strong magnetic fields [8]. When the spin carriers are conduction electrons rather than magnons, the spin current generation due to temperature gradient, which is called the spin-dependent Seebeck effect (SDSE), has been theoretically proposed by Liu and Xie [9] and Zhou *et al.* [10] in the presence of spin-orbit interaction (SOI). Besides the Rashba SOI (RSOI) [11] induced by structure inversion asymmetry considered in Refs. [9, 10], which can be tuned by the external gate voltages and asymmetry doping [12], there is another type of SOI induced by bulk inversion asymmetry, the Desselhaus spin-orbit interaction (DSOI) [13]. The interplays between the RSOI and the DSOI can significantly affect the spin transport [14,15] and spin relaxation [16].

[*] To whom correspondence should be addressed. Email: zhoujunzhou@tongji.edu.cn

The generation of spin polarization by utilizing mesoscopic ring in the presence of SOI with [17,18,19] and without magnetic field [14,20,21,22,23,24,25,26] have been widely studied. The magnetic flux induces the geometric phase of wave functions of electrons in the ring, implying that there is a phase difference between the upper arm and the lower one. It induces the oscillation of the conductance which is known as Aharonov-Bohm (AB) effect [27]. The SOI behaves as an electron momentum dependent in-plane effective magnetic field which lifts the spin degeneracy. Both the dynamical and the spin-dependent geometric phases induced by the SOI are called Aharnov-Casher (AC) phase [28, 29, 30, 31]. The interference due to these phases in a mesoscopic ring can be utilized to generate spin polarization.

In this paper, we present a numerical study of the SDSE in a mesoscopic AB ring in the presence of both the RSOI and the DSOI and magnetic flux. The Green's function method and the Landauer-Büttiker formula are used. The spin Seebeck coefficient (SSC) is studied for various Fermi energy, strengths of RSOI and DSOI, magnetic flux quantum, and temperature.

## 2. Model Systems

We consider a one-dimensional (1D) mesoscopic ring in the presence of both RSOI and DSOI coupled with two semi-infinite leads which is shown in Fig. 1. A magnetic field is applied perpendicular to the plane which results in a magnetic flux across the ring.

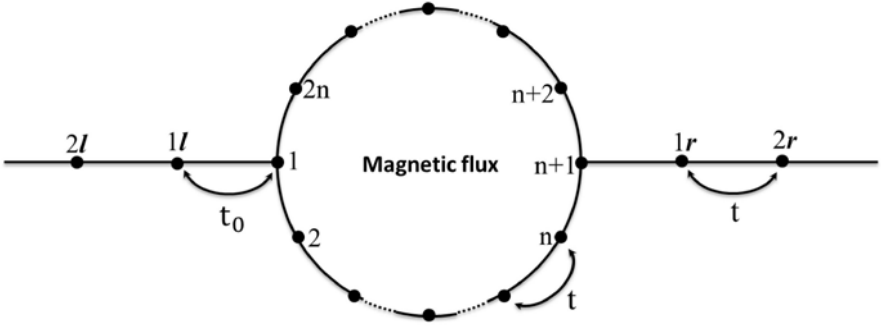

Fig. 1. Schematic of a 1D semiconductor mesoscopic ring attached to two semi-infinite leads.

The tight-binding Hamiltonian of electrons can be written as:

$$H = H_{ring} + H_{lead} + H_i, \qquad (1)$$

where $H_{ring}$ is the Hamiltonian of the isolated ring, $H_{lead}$ is the Hamiltonian of the left ($l$) and right ($r$) semi-infinite leads, and $H_i$ describes the coupling between the ring and leads. They can be written in the nearest-neighboring tight-binding approximations: [14,17]

$$H_{ring} = \sum_i \varepsilon_i \boldsymbol{c}_i^\dagger \boldsymbol{c}_i - \sum_i [(\boldsymbol{c}_i^\dagger \boldsymbol{\tau}_i e^{i\alpha} \boldsymbol{c}_{i+1}) + \text{H.C.}], \qquad (2a)$$

$$H_{lead} = -t \sum_{p=l,r} \sum_j \boldsymbol{d}_{(j+1)p}^\dagger \boldsymbol{d}_{jp} + \text{H.C.}, \qquad (2b)$$

$$H_i = -t_0 (\boldsymbol{d}_{1l}^\dagger \boldsymbol{c}_1 + \boldsymbol{d}_{1r}^\dagger \boldsymbol{c}_{n+1}) + \text{H.C.}, \qquad (2c)$$

here $\boldsymbol{c}_i^\dagger = (c_{i\uparrow}^\dagger, c_{i\downarrow}^\dagger)$ and $\boldsymbol{c}_i = (c_{i\uparrow}, c_{i\downarrow})^T$ are two component creation and annihilation operators of electrons at site $i$ ($i=1, 2,...2n$) on the ring. Similarly, $\boldsymbol{d}_{jp}^\dagger = (d_{jp\uparrow}^\dagger, d_{jp\downarrow}^\dagger)$ and $\boldsymbol{d}_{jp} = (d_{jp\uparrow}, d_{jp\downarrow})^T$ are the two component creation and annihilation operators at the site $j$ ($j=1, 2...\infty$) of lead $p$ ($p=l, r$). $\varepsilon_i$ is the on-site disorder energy strength at site $i$ which is chosen to be zero unless specified. $t$ is the hopping constant between the nearest-neighboring sites. $t_0$ is the hopping constant between the leads and ring. The spin-dependent $2 \times 2$ hopping matrix in Eq. (2a) is written as [32, 33]:

$$\boldsymbol{\tau}_i = t\sigma_0 + i(t^R\cos\theta_i - t^D\sin\theta_i)\sigma_x + i(t^R\sin\theta_i - t^D\cos\theta_i)\sigma_y, \tag{3}$$

where $\sigma_0$ is identity matrix, $t^R$ and $t^D$ are the strengths of RSOI and DSOI, $\sigma_x$ and $\sigma_y$ are the Pauli matrices and $\theta_i = \pi + \frac{\pi}{n}(i - 1/2)$. The Peierls phase factor $e^{i\alpha}$ with $\alpha = (\pi/n)\Phi/\Phi_0$ in Eq. (2a) describes the effect of the magnetic flux, where $\Phi_0 = hc/e$ is the flux quantum.

The spin-dependent transmission coefficient can be estimated with the Green function method. The retarded Green's function can be expressed as:

$$G^R = (E - H - \Sigma_l - \Sigma_r + i\eta)^{-1}, \tag{4}$$

where $E$ is the total energy, $\eta$ is an infinitesimal positive number, $\Sigma_l$ and $\Sigma_r$ are the retarded self-energies for left and right leads, respectively. The spin-dependent transmission coefficient can be expressed as:

$$\mathcal{T}_{\sigma,\sigma'}(E) = \mathrm{T_r}\left[\Gamma_r^\sigma G^R \Gamma_l^{\sigma'} G^A\right], \tag{5}$$

where $\Gamma_p = i[\Sigma_p - \Sigma_p^\dagger] = 2t\sin(k_0 a)$ (with $p=l, r$), $G^A = [G^R]^\dagger$ is the advanced Green's function and $\mathrm{T_r}$ stands for trace.

The electric current across the system is given by $J_e = -e(I_\uparrow + I_\downarrow)$ and the spin current is $J_s = (\hbar/2)(I_\uparrow - I_\downarrow)$. Here, $I_\uparrow$ and $I_\downarrow$ denote the particle current for the spin-up and -down electrons, respectively. They can be obtained by the Landauer-Büttiker formula [34, 35]:

$$I_\sigma = \left(\frac{1}{h}\right)\int dE \mathcal{T}_\sigma(E)[f_l(E) - f_r(E)], \tag{6}$$

where $\mathcal{T}_\sigma(E)$ is defined as $\mathcal{T}_\sigma(E) = \mathcal{T}_{\sigma,\uparrow}(E) + \mathcal{T}_{\sigma,\downarrow}(E)$, $f_l(E)$ and $f_r(E)$ are the electron distribution functions in the left and right leads, respectively. $h$ is the Planck constant. By expanding the distribution function to the first order under low temperature differences and small voltage bias, we obtain the formula for particle current driven by both temperature bias $(T_l - T_r)$ and voltage bias $(V_l - V_r)$ in the linear-response regime [34]:

$$I_\sigma = \left(\frac{e}{h}\right) \int dE \mathcal{T}_\sigma(E) \left(-\frac{\partial f_0}{\partial E}\right) \left[\frac{E - E_F}{eT}(T_l - T_r) - (V_l - V_r)\right], \qquad (7)$$

where $E_F$ is the Fermi energy, $e$ is the electronic charge, $f_0$ is the Fermi distribution. In the case of zero biases, $V_l = V_r$, the SSC defined as $S_s = J_s/(T_l - T_r)$ [10] can be obtained as follows:

$$S_s = \frac{1}{4\pi} \int dE \left(-\frac{\partial f_0}{\partial E}\right)\left(\frac{E - E_F}{T}\right) \Delta \mathcal{T}(E), \qquad (8)$$

where $\Delta \mathcal{T}(E) = \mathcal{T}_\uparrow(E) - \mathcal{T}_\downarrow(E)$, and $T = (T_r + T_l)/2$. In the low temperature limit, Eq. (8) yields by using the Sommerfeld expansion [36, 37]

$$\lim_{T\to 0} S_s = \left[\frac{\pi k_B^2}{12} \frac{d\Delta \mathcal{T}(E)}{dE}\bigg|_{E=E_F}\right] T. \qquad (9)$$

From Eq. (9), we find that SSC depends on temperature linearly at low temperature while the proportionality is determined by the slope of $\Delta \mathcal{T}(E)$.

## 3. Results and discussions

We study the spin-dependent transmission coefficient $\mathcal{T}_\sigma(E)$ and the corresponding SSC of the system by varying four parameters: magnetic flux $\Phi$, Fermi energy $E_F$, the strengths of the RSOI $t^R$ and the DSOI $t^D$. We normalize all the energies by the hopping constant $t$, which is given by $t = \hbar^2/(2m^*a^2)$. For etched InGaAs/GaAs materials [38], the effective mass is $m^* = 0.063 m_e$, where $m_e$ is the free electron mass. Typical value of $a = 10$nm is used as the step in the finite difference calculations. Then $t$=6meV is used throughout the paper. We also set 2n = 100 and $\varepsilon_i = 0$ for a uniform clean ring. To simplify our investigation, we just consider the most simplest case by setting $t_0 = t$ throughout the paper although the coupling between the leads and the conductor is very important for quantum transport [39,40,41].

By analyzing the symmetry of the Hamiltonian of the system, we find that the Hamiltonian in Eq. (2) is invariant under transformations $\sigma_x \to -\sigma_x, \sigma_y \to -\sigma_y, x \leftrightarrow y$, and $t^R \leftrightarrow t^D$. Such invariance implies that the exchange between $t^R$ and $t^D$ leads to $\mathcal{T}_{\sigma,\sigma'}(t^R, t^D, E) = \mathcal{T}_{\bar\sigma,\bar{\sigma'}}(t^D, t^R, E)$ [17], where $\bar\sigma$ means spin reversing.

Therefore, we have $\mathcal{T}_\sigma(t^R, t^D, E) = \mathcal{T}_{\bar{\sigma}}(t^D, t^R, E)$ which can simplify our investigation. This relation further implies that $\Delta\mathcal{T} = 0$ when $t^R = t^D$.

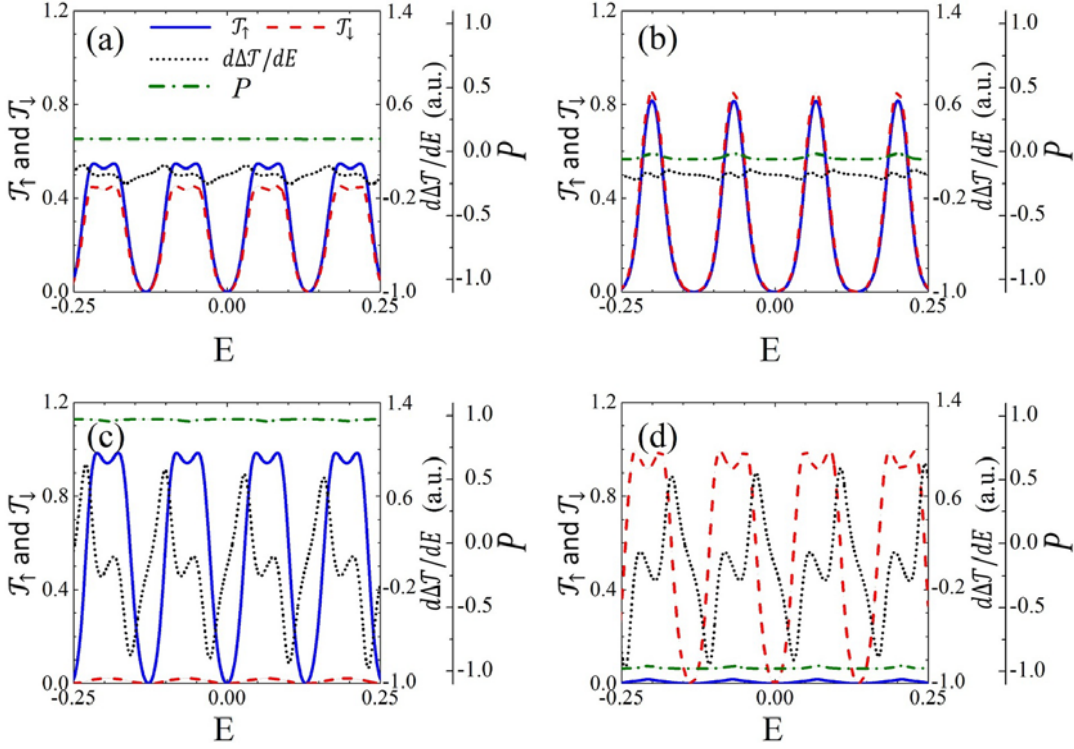

Fig. 2. $\mathcal{T}_\uparrow$ (blue solid curves), $\mathcal{T}_\downarrow$ (red dashed curves), $d\Delta\mathcal{T}/dE$ (black dotted curves), and spin polarization coefficient $P$ (green chain curves) versus the electron energy for different combination of the magnetic fluxes and the strengths of RSOI and DSOI. (a) $\Phi = 0.1\Phi_0, t^D = 0, t^R = 0.33$, (b) $\Phi = 0.1\Phi_0, t^D = 0.32, t^R = 0.07$, (c) $\Phi = 0.4\Phi_0, t^D = 0.16, t^R = 0.13$, and (d) $\Phi = 0.4\Phi_0, t^D = 0.33, t^R = 0.29$.

Fig. 2 shows the spin-up and spin-down transmission coefficients, the slope of their difference, $d\Delta\mathcal{T}/dE$, and the spin polarization $P = (\mathcal{T}_\uparrow - \mathcal{T}_\downarrow)/(\mathcal{T}_\uparrow + \mathcal{T}_\downarrow)$ as functions of the electron energy for various magnetic fluxes and strengths of RSOI and DSOI. We first investigate the cases that $t^D$ is significantly different from $t^R$, i. e. $t^D = 0, t^R = 0.33$ in Fig. 2(a) and $t^D = 0.32, t^R = 0.07$ in Fig. 2(b), while $\Phi = 0.1\Phi_0$ in both figures. The difference between $\mathcal{T}_\uparrow$ and $\mathcal{T}_\downarrow$ is tiny which leads to small $d\Delta\mathcal{T}/dE$ and spin polarization. In contrast, we studied the cases that $t^D$ is close to (but not equals to) $t^R$, for example $t^D = 0.16, t^R = 0.13$ in Fig. 2(c) and $t^D = 0.33, t^R = 0.29$ in Fig. 2(d), when $\Phi = 0.4\Phi_0$. Due to different quantum interference, the spin-down electron is strongly reflected and the spin-up electron can transmit through the ring for most energies as shown in Fig. 2(c). An opposite feature is shown in Fig. 2(d). Therefore, a large positive spin polarization close to 1 is obtained in Fig. 2(c) and a negative one close to -1 is obtained in Fig. 2(d). The strongly energy-dependent transmission coefficients and remarkable difference between $\mathcal{T}_\uparrow$ and $\mathcal{T}_\downarrow$ in Fig. 2(c) and Fig. 2(d) results in a large $d\Delta\mathcal{T}/dE$ which also oscillate with electron energy. According to Eqs. (8) and (9), such large $d\Delta\mathcal{T}/dE$ is preferred to achieve significant SDSE.

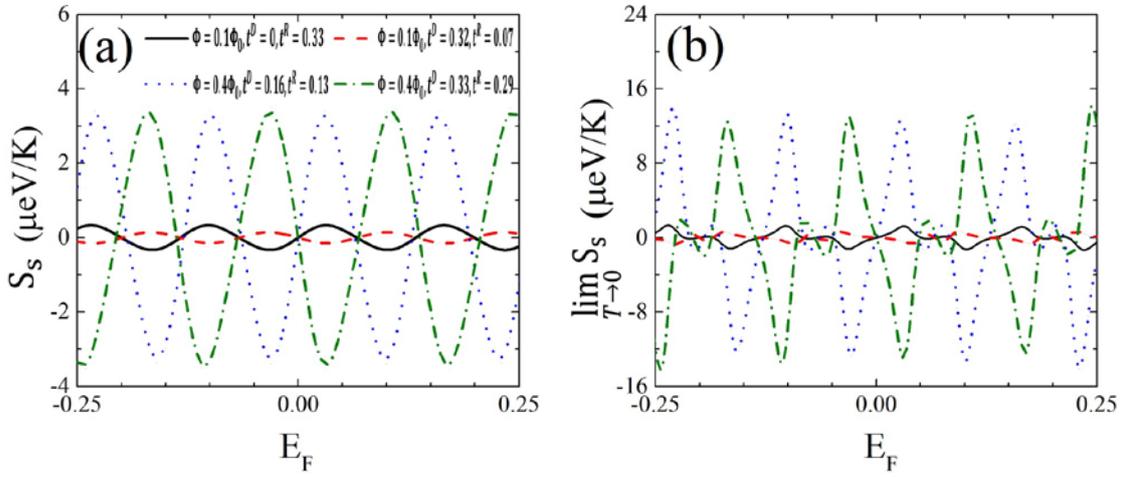

Fig.3. (a) $S_s$ calculated from Eq. (8) and (b) $\lim_{T\to 0} S_s$ calculated from Eq. (9) versus the Fermi energy with the same parameters as in Fig. 2. The temperature is $T=1K$.

Fig. 3(a) shows the Fermi energy dependence of SSC calculated from Eq. (8) when $T=1K$ for the same cases as in Fig. 2. The SSC for all cases strongly oscillate with Fermi energy and both positive and negative values can be found. The largest absolutely value of SSC could be $3.5 \mu eV/K$. It is obvious that $S_s$ for $\Phi = 0.4\Phi_0, t^D = 0.16, t^R = 0.13$ and $\Phi = 0.4\Phi_0, t^D = 0.33, t^R = 0.29$ are much larger than $S_s$ for the other two cases, because the slopes of $\Delta\mathcal{T}$ for the latter two cases are larger than that for the former two cases as shown in Fig. 2. In the low-temperature regime, the SSC in Eq. (8) can be approximated by Eq. (9), which shows that SSC is proportional to the temperature $T$. To study the temperature effect, Fig. 3(b) shows the SSC calculated from Eq. (9), i. e. $\lim_{T\to 0} S_s$, with exact the same parameters. Comparing with Fig. 3(a), we find that SSC calculated from Eq. (9) is larger than the SSC calculated from Eq. (8) at 1K. Since the temperature would smear the sharp slope of $\Delta \mathcal{T}$, smoother curves of SSC are observed. Through systematic calculations, which are not completely shown in this paper, we find that the Eq. (9) is a good approximation only when *T<0.1K*. Above 0.1K, Eq. (9) gives an overestimated value.

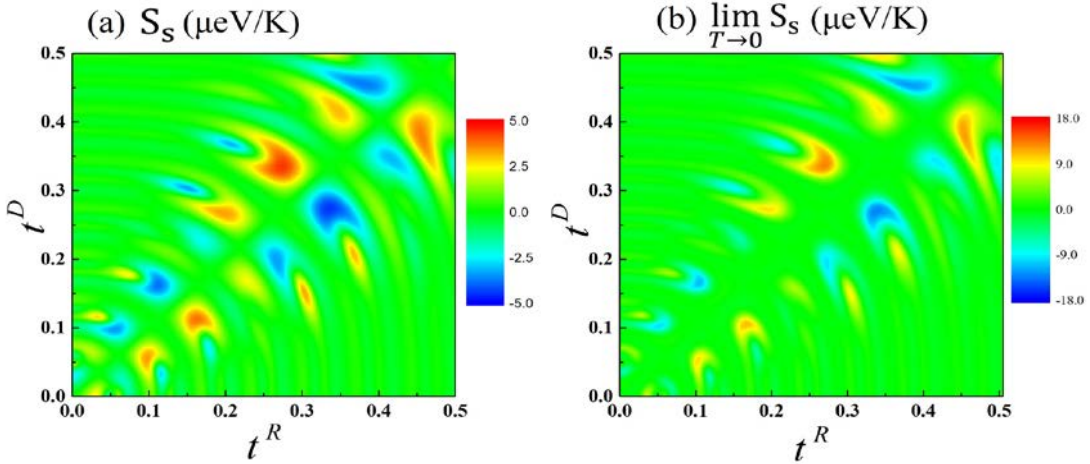

Fig. 4 (a) SSC calculated from Eq. (8) and (b) $\lim_{T\to 0} S_s$ calculated from Eq. (9) versus the strengths of RSOI and DSOI when $\Phi = 0.1\Phi_0$, $E_F = 0.168t$ and *T=0.8K*.

Fig. 4(a) shows the SSC calculated from Eq. (8) as functions of the strengths of both RSOI and DSOI when $\Phi = 0.1\Phi_0$, $E_F = 0.168t$ and *T=0.8K*. We find that $S_S$ is anti-symmetric along the diagonal line, in other words the sign of $S_S$ can be reversed when $t^D \leftrightarrow t^R$, because of the relation $\mathcal{T}_\sigma(t^R, t^D, E) = \mathcal{T}_{\bar{\sigma}}(t^D, t^R, E)$ we discussed above. Consequently, zero $S_S$ is found when $t^D = t^R$. The maximum value $S_S$ can be found when $|t^R - t^D| \in 0.04 \sim 0.08$. For instance, the largest SSC are found to be $4.42 \mu eV/K$ ($-4.42 \mu eV/K$) when $t^R = 0.27$ (0.34) and $t^D = 0.34$ (0.27). The existence of maximum $S_S$ can be explained as follows. Both $\mathcal{T}_\uparrow$ and $\mathcal{T}_\downarrow$ vanish for strong RSOI and DSOI because of the effective periodic potential [42] which leads to $\Delta \mathcal{T} = 0$ and $S_S = 0$. Zero SSC can also be obtained when $t^D = t^R$ due to spin degeneracy [42]. Therefore, in between these two cases, $\Delta \mathcal{T}$, which is between -1 and 1, must be continuous. Then the energy dependent $\Delta \mathcal{T}$ as shown in Fig. 2 leads to an oscillation of SSC between positive maximum and negative minimum. In Fig. 4(b), we also show $\lim_{T\to 0} S_S$ calculated from Eq. (9) for comparison. Overestimation of $S_S$ is induced by the temperature smearing effect. For example, $S_S = 13.2\mu eV/K$ ($-13.2\mu eV/K$) when $t^R = 0.27$ (0.34) and $t^D = 0.34$ (0.27), which are much larger than the values in Fig. 4(a), are obtained. The findings in Fig. 4 provide a possible way to estimate the strength of the DSOI by searching for exact zero SSC, which is independent on Fermi energy when $t^D = t^R$. One can tune the strength of the RSOI by the external electric fields.

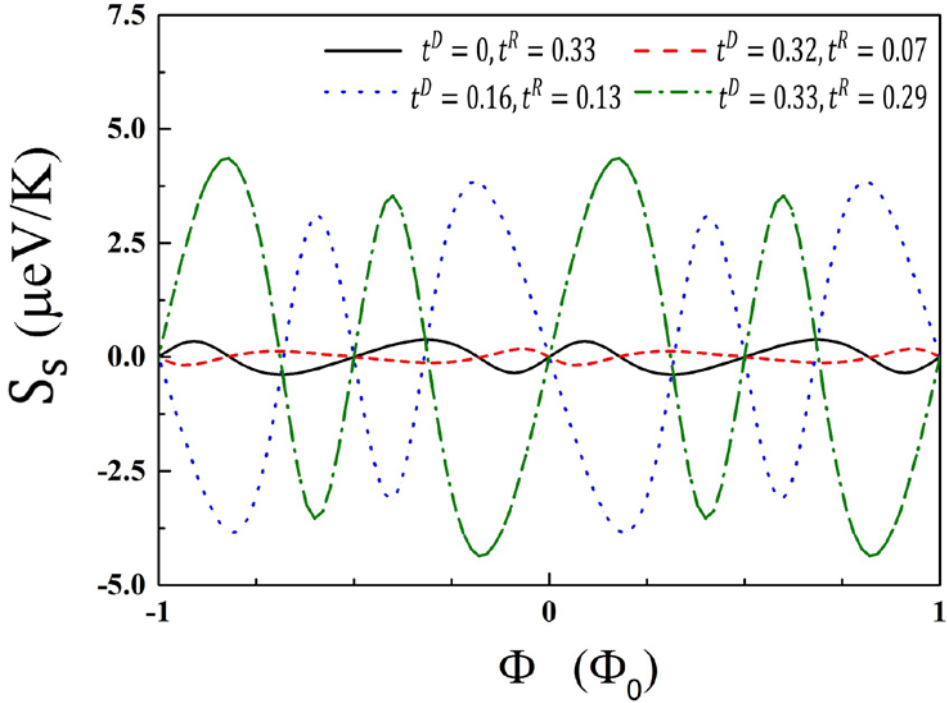

Fig. 5. SSC versus magnetic flux for fixed values for four combinations of strengths of RSOI and DSOI when $E_F = 0.168t$ and $T = 1K$.

Fig. 5 shows the SSC as a function of the magnetic flux for different strengths of RSOI and DSOI when $E_F = 0.168t$ and $T = 1K$. A perfect periodicity of $S_S$ can be found since the Hamiltonian is a periodic function, i. e. $H_{ring}(\Phi) = H_{ring}(\Phi + N\Phi_0)$ where $N$ is arbitrary integer. The maximum value of SSC is $4.4\mu eV/K$ when $\Phi = (N + 0.18)\Phi_0$ and $-4.4\mu eV/K$ when $\Phi = (N + 0.82)\Phi_0$ for $t^R = 0.29$ and $t^D = 0.33$. We also find that $S_S$ is an odd function of magnetic flux around $(N/2)\Phi_0$. The reason is that $H_{ring}^{\dagger}\left(\frac{N\Phi_0}{2}-\Phi\right)=H_{ring}\left(-\frac{N\Phi_0}{2}+\Phi\right)=H_{ring}(\frac{N\Phi_0}{2}+\Phi)$ where the periodicity of Hamiltonian is used. This relation leads to $\mathcal{T}_\sigma(\frac{N\Phi_0}{2}-\Phi)=\mathcal{T}_{\bar{\sigma}}(\frac{N\Phi_0}{2}+\Phi)$ which yields $S_S\left(\frac{N\Phi_0}{2}-\Phi\right)=-S_S\left(\frac{N\Phi_0}{2}+\Phi\right)$. This feature results in a zero SSC when the magnetic flux is half-integer times of magnetic flux quantum.

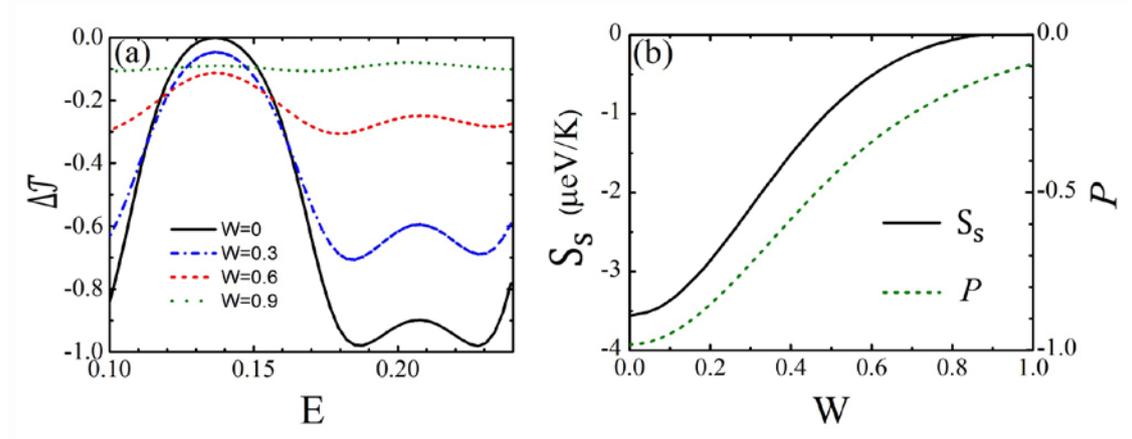

Fig. 6. (a) $\Delta\mathcal{T}$ versus the electron energy for different disorder strengths. (b) SSC and spin polarization versus the disorder strength. $t^D=0.33, t^R=0.29$, $\Phi=0.4\Phi_0$, $E_F=0.168t$ and $T$=1K are used in the calculations.

Finally, we discuss the effect of disorder on the SSC and the spin polarization as shown in Fig. 6. The disorder is introduced by choosing $\varepsilon_i$ to be random values in the range $[-W/2, W/2]$ where W is the strength of disorder. The numerical results are obtained by averaging over 5000 random disorder configurations. Fig. 6(a) shows that the difference between $\mathcal{T}_\uparrow$ and $\mathcal{T}_\downarrow$ becomes smaller for most electron energies and the variation becomes smoother for larger W. Therefore, the spin polarization and the SSC decrease with the increase of the disorder strength as shown in Fig. 6(b). The SSC vanishes when W is larger than 0.85.

## 4. Summary

In summary, we have studied the spin-dependent Seebeck effect in an Aharonov-Bohm ring in the presence of both Rashba and Dresselhaus spin-orbit interaction. The spin Seebeck coefficient strongly depends on the Fermi energy, strengths of RSOI and DSOI, magnetic flux, and temperature. Maximum value of spin Seebeck coefficient can be found when the strengths of RSOI and DSOI are slightly different from each other while the magnetic flux is not half-integer times of magnetic quantum. Moreover, increasing temperature leads to a reduction of spin Seebeck coefficient due to smearing effect. The spin Seebeck coefficient can be killed by strong disorder.

Y. Li acknowledge the support from Tongji University under grant NO. 2013KJ025. J. Z. is also supported by the program for New Century Excellent Talents in Universities (Grant No. NCET-13-0431) and the Program for Professor of Special Appointment (Eastern Scholar) at Shanghai Institutions of Higher Learning (Grant No. TP2014012). B. Liu would like to thank Ms. T. Y. Lu for help.